\documentstyle[preprint,tighten,eqsecnum,aps,floats,amssymb]{revtex}

\begin{document}
\draft

\title{
High-order perturbative expansions of multi-parameter $\Phi^4$ quantum
field theories
}
\author{Andrea Pelissetto,${}^1$  Ettore Vicari${}^2$ }
\address{$^1$ Dipartimento di Fisica dell'Universit\`a di Roma 
``La Sapienza" and INFN,\\ 
P.le Aldo Moro 2, I-00185 Roma, Italy}
\address{$^2$
Dipartimento di Fisica dell'Universit\`a di Pisa 
and INFN, \\ 
Largo Pontecorvo 2, I-56127 Pisa, Italy
}
\address{
{\bf e-mail: 
{\tt Andrea.Pelissetto@roma1.infn.it},
{\tt Ettore.Vicari@df.unipi.it}
}}

\maketitle

\begin{abstract}
  
  We present high-order pertubative expansions of multi-parameter $\Phi^4$
  quantum field theories with an $N$-component fundamental field, containing
  up to 4th-order polynomials of the field.  Multi-parameter $\Phi^4$ theories
  generalize the simplest $O(N)$-symmetric $\Phi^4$ theories, and describe
  more complicated symmetry breaking patterns.  These notes collect several
  high-order perturbative series of physically interesting multi-parameter
  $\Phi^4$ theories, to five or six loops.  We consider the
  O$(M)\otimes$O$(N)$-symmetric $\Phi^4$ model, the so-called $MN$ model, and
  a spin-density-wave $\Phi^4$ model containing five quartic terms.  The
  corresponding Tables of the coefficients are reported in Ref.~\cite{epaps}.

\end{abstract}

\newpage


\section{introduction}

In the framework of the renormalization-group (RG) approach to critical
phenomena, a quantitative description of many continuous phase transitions can
be obtained by considering an effective Landau-Ginzburg-Wilson (LGW) $\Phi^4$
field theory, containing up to fourth-order powers of the field components.
The simplest example is the O($N$)-symmetric $\Phi^4$ theory, defined by the
Lagrangian density
\begin{equation}
{\cal L}_{O(N)} =  
{1\over 2} \sum_i (\partial_\mu \Phi_{i})^2 + 
{1\over 2} r \sum_i \Phi_{i}^2  + 
{1\over 4!} u (\sum_{i} \Phi_i^2 )^2
\label{HON}
\end{equation}
where $\Phi$ is an $N$-component real field.  These $\Phi^4$ theories describe
are characterized by the symmetry breaking O($N$)$\rightarrow$O($N-1$).  See,
e.g., Refs.~\cite{ZJ-book,PV-rev} for recent reviews discussing these models.
Beside the transitions described by O($N$) models, there are also other
physically interesting transitions described by more general
Landau-Ginzburg-Wilson (LGW) $\Phi^4$ field theories, characterized by more
complex symmetries and symmetry breaking patterns.  The general LGW $\Phi^4$
theory for an $N$-component field $\Phi_i$ can be written as
\begin{equation}
{\cal L} = 
{1\over 2} \sum_i (\partial_\mu \Phi_{i})^2 + 
{1\over 2} \sum_i r_i \Phi_{i}^2  + 
{1\over 4!} \sum_{ijkl} u_{ijkl} \; \Phi_i\Phi_j\Phi_k\Phi_l  
\label{generalH}
\end{equation}
where the number of independent parameters $r_i$ and $u_{ijkl}$ depends on the
symmetry group of the theory.  Here, we are only assuming a parity symmetry
which forbids third-order terms.  
An interesting class of models are those in which $\sum_i \Phi^2_i$ is the
unique quadratic polynomial invariant under the symmetry group of the theory,
corresponding to the case all field components become critical simultaneously.
This requires that all $r_i$ are equal, $r_i = r$, and $u_{ijkl}$ must be such
not to generate other quadratic invariant terms under RG transformations, for
example, it must satisfy the trace condition~\cite{ZJ-book} $\sum_i u_{iikl}
\propto \delta_{kl}$.  
In these models, criticality is driven by tuning the single parameter $r$,
which physically may correspond to the reduced temperature.

More general LGW $\Phi^4$ theories, which allow for the presence of
independent quadratic parameters $r_i$, must be considered to describe
multicritical behaviors where there are independent correlation lengths that
diverge simultaneously, which may arise from the competition of distinct types
of ordering, see e.g. Ref.~\cite{CPV-03} and references therein.  Note that,
like the simplest O($N$) models, all multi-parameter $\Phi^4$ field theories
are expected to be trivial in four dimensions.

High-order perturbative expansions, to five and six loops, of several
multi-parameter $\Phi^4$ theories have been computed in
Refs.~\cite{PV-rev,CPV-03,KS-95,PS-00,CPV-00,PV-00,PRV-01,PRV-01-2,Parruccini-03,CP-03,BPV-03,DPV-04,CPPV-04,CP-04,CPV-05,PV-05,BPV-05,DPV-06}.
These notes collect several high-order series of physically interesting
multi-parameter $\Phi^4$ theories, to five or six loops.  The corresponding
Tables of the coefficients are reported in Ref.~\cite{epaps}.  We consider two
perturbative schemes: the three-dimensional (3D) massive zero-momentum (MZM)
scheme in three dimensions and the massless (critical) $\overline{\rm MS}$
scheme.  In the MZM scheme, one expands in powers of the MZM quartic couplings
$g_{ijkl}$, defined by
\begin{eqnarray}
\Gamma^{(2)}_{ij}(p) = \delta_{ij} Z_\phi^{-1} \left[ m^2+p^2+O(p^4)\right],
\qquad
\Gamma^{(4)}_{ijkl}(0) = m\,Z_\phi^{-2} \,g_{ijkl}
\label{mzmcom}
\end{eqnarray}
The $\overline{\rm MS}$ scheme is based on a minimal subtraction procedure
within the dimensional regularization, and can give rise to an $\epsilon\equiv
4-d$ expansion, and also 3D expansions in the renormalized $\overline{\rm MS}$
couplings $g_{ijkl}$ by setting $\epsilon=1$ after renormalization.  The RG
flow is determined by the FPs, which are common zeroes $g^*_{ijkl}$ of the
$\beta$-functions, $\beta_{ijkl}(g_{abcd})\equiv m {\partial g_{ijkl}/\partial
  m}$ and $\beta_{ijkl}(g_{abcd})\equiv \mu {\partial g_{ijkl}/\partial \mu}$
in the MZM and $\overline{\rm MS}$ schemes respectively.  We report series for
the O$(M)\otimes$O$(N)$-symmetric $\Phi^4$ model, the so-called $MN$ model,
and a spin-density-wave $\Phi^4$ model containing five quartic terms. We also
mention that high-order perturbative series for $U(N)\times U(N)$,
SU(4)$\otimes$SU(4), U($N$) and SU($N$) $\Phi^4$ field theories have been
computed in Refs.~\cite{BPV-03,CP-04,BPV-05}.

\section{The O$(M)\otimes$O$(N)$-symmetric model}
\label{secomn}

The O($M)\otimes$O($N)$-symmetric $\Phi^4$ model is defined by the
Hamiltonian density
\begin{eqnarray}
  {1\over2}
\sum_{ai} \Big[ \sum_\mu (\partial_\mu \Phi_{ai})^2 + r \Phi_{ai}^2 \Big]   
+ {u_0\over 4!}  \Big( \sum_{ai} \Phi_{ai}^2 \Big)^2 
+ {v_0\over 4!}  \Big[ \sum_{i,j} 
\Big( \sum_a \Phi_{ai} \Phi_{aj}\Big)^2 - \Big(\sum_{ai} \Phi_{ai}^2\Big)^2 \Big],
\label{Hch}
\end{eqnarray}
where $\Phi_{ai}$ is a real $N\times M$ matrix field ($a=1,\ldots,N$
and $i=1,\ldots,M$).

We also consider the four independent quadratic perturbations $Q^{(k)}$ that break
the O($M$)$\otimes$O($N$) symmetry, i.e.  
\begin{eqnarray}
Q^{(1)}_{aibj} &=& \Phi_{ai}\Phi_{bj} -\Phi_{aj}\Phi_{bi},  \label{operators} \\
Q^{(2)}_{aibj} &=& \case{1}{2} \left( \Phi_{ai}\Phi_{bj}+\Phi_{aj}\Phi_{bi}\right)
- \case{1}{M} \delta_{ab} \Phi_{ci}\Phi_{cj} - \case{1}{N} \delta_{ij} \Phi_{ak}\Phi_{bk} 
+ \case{1}{MN} \delta_{ab} \delta_{ij} \Phi_{ck}\Phi_{ck}, \nonumber \\
Q^{(3)}_{ij} &=& \Phi_{ci}\Phi_{cj} 
- \case{1}{N} \delta_{ij} \Phi_{ck}\Phi_{ck}, \nonumber\\
Q^{(4)}_{ab} &=& \Phi_{ak}\Phi_{bk} 
- \case{1}{M} \delta_{ab} \Phi_{ck}\Phi_{ck}.
\nonumber
\end{eqnarray}
The above four perturbations are related to different representations
of the O($M$) and O($N$) groups. Therefore, they do no mix under
renormalization-group (RG) transformations. 

In the following we report the perturbative expansions in the massive
zero-momentum (MZM) scheme and in the minimal-subtraction ($\overline {\rm
  MS}$) scheme.  For further details see
Refs.~\cite{PV-rev,PRV-01,PRV-01-2,Parruccini-03,CP-03,DPV-04,CPPV-04,CPV-05}.

\subsection{The 3D massive zero-momentum perturbative expansion}
\label{mzm}

In the MZM scheme the theory is renormalized
by introducing a set of zero-momentum conditions for the one-particle
irreducible two-point and four-point correlation functions:
\begin{equation}
\Gamma^{(2)}_{ai,bj}(p) = 
\delta_{ai,bj} Z_\phi^{-1} \left[ m^2+p^2+O(p^4)\right],
\label{ren1}  
\end{equation}
where $\delta_{ai,bj} \equiv \delta_{ab}\delta_{ij}$,
\begin{equation}
\Gamma^{(4)}_{ai,bj,ck,dl}(0) = 
Z_\phi^{-2} m \left( u S_{ai,bj,ck,dl} +  v C_{ai,bj,ck,dl} \right),
\label{ren2}  
\end{equation}
and $S$, $C$ are appropriate tensorial factors associated with
the two quartic terms of Hamiltonian (\ref{Hch}):
\begin{eqnarray}
S_{ai,bj,ck,dl} \equiv && \case{1}{3}
(\delta_{ai,bj}\delta_{ck,dl} + \delta_{ai,ck}\delta_{bj,dl} 
+ \delta_{ai,dl}\delta_{bj,ck} ), \label{tensors}
\\
C_{ai,bj,ck,dl} \equiv && \case{1}{6} \left[
\delta_{ab}\delta_{cd} ( \delta_{ik}\delta_{jl} + \delta_{il}\delta_{jk})
+ \delta_{ac}\delta_{bd} ( \delta_{ij}\delta_{kl} + \delta_{il}\delta_{jk} ) + 
\delta_{ad}\delta_{bc} ( \delta_{ij}\delta_{kl} + \delta_{ik}\delta_{jl} ) 
\right]- S_{ai,bj,ck,dl}  \nonumber
\end{eqnarray}
Eqs.~(\ref{ren1}) and (\ref{ren2}) relate the mass scale 
(inverse correlation length) $m$,
and the zero-momentum quartic couplings $u$ and $v$ to the
corresponding Hamiltonian parameters $r$, $u_0$, and $v_0$.  In
addition, the function $Z_t$ is defined by the relation
\begin{equation}
\Gamma^{(1,2)}_{ai,bj}(0) = \delta_{ai,bj} Z_t^{-1}, 
\label{ren3}
\end{equation}
where $\Gamma^{(1,2)}$ is the one-particle irreducible two-point function
with an insertion of $\case{1}{2}\Phi^2$.  The RG flow in the
quartic-coupling $u,v$ plane is determined by the $\beta$-functions
\begin{eqnarray}
\beta_u(u,v) = \left. m{\partial u\over \partial m} \right|_{u_0,v_0},\qquad
\beta_v(u,v) = \left. m{\partial v\over \partial m} \right|_{u_0,v_0}.
\label{betamzm}
\end{eqnarray}
Their common zeroes are the fixed points (FP's) of the RG flow.
The RG functions $\eta_\phi$ and $\eta_t$ 
associated with the standard critical exponents are
\begin{eqnarray}
\eta_{\phi,t} (u,v) = {\partial \ln Z_{\phi,t} \over \partial \ln m}
= \beta_u {\partial \ln Z_{\phi,t} \over \partial u} +
\beta_v {\partial \ln Z_{\phi,t} \over \partial v} ,
\label{etaphitmzm}
\end{eqnarray}
In order to compute the RG dimensions of the quadratic operators
(\ref{operators}), one computes the four functions $Z_{Qk}$ defined
through the relation $\Gamma^{(Qk,2)}_{ai,bj}(0) = q_{k,ai,bj}
Z_{Qk}^{-1}$, where $\Gamma^{(Qk,2)}$ is the one-particle irreducible
two-point function with an insertion of $Q^{(k)}$ 
($q_{k,ai,bj}$ are the appropriate tensorial factors). One then derives
the corresponding RG functions $\eta_{Qk}$ through the equation
\begin{eqnarray}
\eta_{Qk} (u,v) = {\partial \ln Z_{Qk} \over \partial \ln m}
\end{eqnarray}
The critical exponents $\eta$ and $\nu$ and the RG dimensions $y_{Qk}$ of the
quadratic operators $Q^{(k)}$ are obtained by
evaluating the RG functions at the fixed point $u^*$, $v^*$, as 
\begin{eqnarray}
&&\eta = \eta_\phi(u^*,v^*), \label{exponents} \\
&&\nu  = \left[ 2 - \eta_\phi(u^*,v^*) + \eta_t(u^*,v^*)\right]
^{-1},\nonumber \\
&&y_{Qk}  = 2 - \eta_\phi(u^*,v^*) + \eta_{Qk}(u^*,v^*). \nonumber
\end{eqnarray}

The 3D expansions of the RG functions in powers of the zero-momentum
couplings $u,v$ are [we set $A=3/(16\pi)$]
\begin{eqnarray}
\beta_{u}  = \sum_{ijkl} b^{(u)}_{ijkl} M^k N^l u^i v^j 
= 
&&
- u +  \case{8 + MN}{9} A u^2 
- \case{2(1-M)(1-N)}{9} A u v 
+ \case{(1-M)(1-N)}{9} A v^2 
\\&&
- \case{760  + 164 MN}{2187 } A^2 u^3  
+\case{400(1-M)(1-N)}{2187} A^2 u^2 v
-\case{ 118 (1-M)(1-N)}{729} A^2 u v^2
\nonumber\\ 
&&+\case{90 (1-M)(1-N)}{2187} A^2 v^3 + ...
\nonumber\\
\beta_{v}  =
\sum_{ijkl} b^{(v)}_{ijkl} M^k N^l u^i v^j =
&&- v - \case{8-M-N}{9} A v^2 + \case{4}{3} A u v   -
\case{1480 + 92 MN}{2187 } A^2 u^2 v 
\nonumber\\
&& + \case{1912 - 400 M - 400 N + 184 MN}{2187 } A^2 u v^2
- \case{298  - 106 M - 106 N + 40 MN}{729} A^2 v^3 + ...
\nonumber \\
\eta_\phi = \sum_{ijkl} e^{(\phi)}_{ijkl} M^k N^l u^i v^j  = 
&&\case{16 + 8 MN}{2187} A^2 u^2 
- \case{16(1-M)(1-N)}{2187} A^2 u v 
+\case{4(1-M)(1-N)}{729} A^2 v^2 + ...
\nonumber\\
\eta_t = \sum_{ijkl} e^{(t)}_{ijkl} M^k N^l u^i v^j 
= &&-\case{2+MN}{9} A u  + \case{(1-M)(1-N)}{9} A v  
+ \case{4 +2 MN}{81} A^2 u^2 
-\case{4(1-M)(1-N)}{81} A^2 u v 
\nonumber \\&&
+\case{(1-M)(1-N)}{27} A^2 v^2 + ...
\nonumber\\
\eta_{Q1} = 
\sum_{ijkl} e^{(Q1)}_{ijkl} M^k N^l u^i v^j = 
&&- \case{2}{9} A u  + \case{1}{9} A v
+ \case{12+2MN}{243} A^2 u^2 
-  \case{28-4M-4N+4MN}{243} A^2 u v 
\nonumber \\&& + 
\case{17 - 5M -5N + 2MN}{243} A^2 v^2 
+ ... \nonumber\\
\eta_{Q2} = 
 \sum_{ijkl} e^{(Q2)}_{ijkl} M^k N^l u^i v^j = 
&&- \case{2}{9} A u  + \case{1}{9} A v
+ \case{12+2MN}{243} A^2 u^2 
-  \case{12-4M-4N+4MN}{243} A^2 u v 
\nonumber \\&&
+ \case{9 - 3M -3N + 2MN}{243} A^2 v^2 
+ ... \nonumber\\
\eta_{Q3} = 
 \sum_{ijkl} e^{(Q3)}_{ijkl} M^k N^l u^i v^j = 
&&- \case{2}{9} A u  + \case{1-N}{9}A v
+ \case{12+2MN}{243} A^2 u^2 
-  \case{12-4M -12N + 4MN}{243} A^2 u v 
\nonumber \\&& 
+ \case{3 - M -3N + MN}{81} A^2 v^2 
+ ... \nonumber\\
\eta_{Q4} = 
 \sum_{ijkl} e^{(Q4)}_{ijkl} M^k N^l u^i v^j = 
&&- \case{2}{9} A u  + \case{1-M}{9}A v
+ \case{12+2MN}{243} A^2 u^2 
-  \case{12-12M -4N + 4MN}{243} A^2 u v 
\nonumber \\&&
+ \case{3 - 3M -N + MN}{81} A^2 v^2 
+ ... \nonumber
\end{eqnarray}
The values of the coefficients $b^{(u)}_{ijkl}$, $b^{(v)}_{ijkl}$,
$e^{(\phi)}_{ijkl}$, $e^{(t)}_{ijkl}$, and $e^{(Qk)}_{ijkl}$ up to six loops
are reported in the file {\tt OMN-MZM.TXT} attached to Ref.~\cite{epaps}.
Each line of this file contains 7 numbers. The first one indicates the
quantity at hand: 1,2,3,4,5,6,7 correspond respectively to $\beta_u$,
$\beta_v$, $\eta_\phi$, $\eta_t$, $\eta_{Q1}$, $\eta_{Q2}$, $\eta_{Q3}$ (the
expansion of $\eta_{Q4}$ can be obtained from that of $\eta_{Q3}$ by
interchanging $M$ with $N$); the second integer number gives the number of
loops; the subsequent four integer numbers are the indices $i,j,k,l$; finally,
the last real number is the value of the coefficient.

\subsection{The minimal-subtraction pertubative expansion}
\label{msbar}

In the $\overline{\rm MS}$ scheme one sets
\begin{eqnarray}
\Phi &=& [Z_\phi(u,v)]^{1/2} \Phi_R, \\
u_0 &=& A_d \mu^\epsilon Z_u(u,v) , \nonumber \\
v_0 &=& A_d \mu^\epsilon Z_v(u,v) , \nonumber
\end{eqnarray}
where the renormalization functions $Z_\phi$, $Z_u$, and $Z_v$ are determined
from the divergent part of the two- and four-point one-particle irreducible
correlation functions computed in dimensional regularization.  They are
normalized so that $Z_\phi(u,v) \approx 1$, $Z_u(u,v) \approx u$, and
$Z_v(u,v) \approx v$ at tree level.  Here $A_d$ is a $d$-dependent constant
given by $A_d\equiv 2^{d-1} \pi^{d/2} \Gamma(d/2)$.  Moreover, one defines a
mass renormalization constant $Z_t(u,v)$ by requiring $Z_t \Gamma^{(1,2)}$ to
be finite when expressed in terms of $u$ and $v$.  Here $\Gamma^{(1,2)}$ is
the one-particle irreducible two-point function with an insertion of
$\case{1}{2}\Phi^2$.  The $\beta$ functions,
\begin{equation}
\beta_u (u,v) = \mu \left. {\partial u \over \partial \mu} \right|_{u_0,v_0},
\qquad
\beta_v (u,v) = \mu \left. {\partial v \over \partial \mu} \right|_{u_0,v_0},
\end{equation}
have a simple dependence on $d$:
\begin{equation}
\beta_u = (d-4) u + B_u(u,v),\qquad
\beta_v = (d-4) v + B_v(u,v),
\label{Bdef}
\end{equation}
where the functions $B_u(u,v)$ and $B_v(u,v)$ are independent of $d$.  The RG
dimensions of the quadratic operators $Q^{(k)}$ are obtained by computing the
renormalization functions $Z_{Qk}(u,v)$. These functions are determined by
requiring $Z_{Qk} \Gamma^{(Qk,2)}$ to be finite when expressed in terms of $u$
and $v$.  Here $\Gamma^{(Qk,2)}$ is the one-particle irreducible two-point
function with an insertion of $Q^{(k)}$.  The RG functions $\eta_\phi$ and
$\eta_t$, associated with the critical exponents, are obtained from
\begin{equation}
\eta_{\phi,t}(u,v) 
=  \left. {\partial \log Z_{\phi,t} \over \partial \log \mu} \right|_{u_0,v_0}.
\end{equation}
The same equation allows the determination of $\eta_{Qk}$ from $Z_{Qk}$.  The
RG functions $\eta_\phi$, $\eta_t$, and $\eta_{Qk}$ are independent of $d$.
The standard critical exponents $\eta$ and $\nu$, and the RG dimensions
$y_{Qk}$ of the quadratic perturbations $Q_k$ are obtained by using
Eq.~(\ref{exponents}).

The expansions of the RG functions in powers of the $\overline{\rm MS}$
couplings $u,v$ are
\begin{eqnarray}
B_u = 
\sum_{ijkl} b^{(u)}_{ijkl} M^k N^l u^i v^j=
&& \case{8+MN}{6} u^2 - \case{(1-M)(1-N)}{3} uv 
+\case{(1-M)(1-N)}{6} v^2
-\case{14+3MN}{12} u^3 
\\ && + 
\case{11 (1-M)(1-N)}{18} u^2 v  
-\case{13 (1-M)(1-N)}{24} u v^2 
+ \case{5(1-M)(1-N)}{36} v^3 + ...
\nonumber\\
B_v =  \sum_{ijkl} b^{(v)}_{ijkl} M^k N^l u^i v^j = 
&&2 u v  - \case{8-M-N}{6} v^2 - \case{82+ 5 MN}{36} u^2 v 
+  \case{53 - 11 M-11 N + 5MN}{18} u v^2
\nonumber\\ &&
- \case{99 - 35 M-35 N + 13MN}{72} v^3
+... \nonumber\\
\eta_\phi=  \sum_{ijkl} e^{(\phi)}_{ijkl} M^k N^l u^i v^j =
&&\case{2+MN}{72} u^2  - \case{(1-M)(1-N)}{36} u v + \case{(1-M)(1-N)}{48} v^2 
+...\nonumber\\
\eta_t = 
 \sum_{ijkl} e^{(t)}_{ijkl} M^k N^l u^i v^j = 
&&- \case{2+MN}{6} u  + \case{(1-M)(1-N)}{6} v 
\nonumber \\
&& 
+ \case{2+MN}{12} u^2 -  \case{(1-M)(1-N)}{6} u v 
+ \case{(1-M)(1-N)}{8} v^2 
+ ... \nonumber\\
\eta_{Q1} = 
 \sum_{ijkl} e^{(Q1)}_{ijkl} M^k N^l u^i v^j = 
&&- \case{1}{3} u  + \case{1}{2} v
+ \case{6+MN}{36} u^2 
-  \case{7-M-N+MN}{18} u v + 
\case{17 - 5M -5N + 2MN}{72} v^2 
+ ... \nonumber\\
\eta_{Q2} = 
 \sum_{ijkl} e^{(Q2)}_{ijkl} M^k N^l u^i v^j = 
&&- \case{1}{3} u  + \case{1}{6} v
+ \case{6+MN}{36} u^2 
-  \case{3-M-N+MN}{18} u v + 
\case{9 - 3M -3N + 2MN}{72} v^2 
+ ... \nonumber\\
\eta_{Q3} = 
 \sum_{ijkl} e^{(Q3)}_{ijkl} M^k N^l u^i v^j = 
&&- \case{1}{3} u  + \case{1-N}{6} v
+ \case{6+MN}{36} u^2 
-  \case{3-M -3N + MN}{18} u v + 
\case{3 - M -3N + MN}{24} v^2 
+ ... \nonumber\\
\eta_{Q4} = 
\sum_{ijkl} e^{(Q4)}_{ijkl} M^k N^l u^i v^j = 
&&- \case{1}{3} u  + \case{1-M}{6} v
+ \case{6+MN}{36} u^2 
-  \case{3-3M -N + MN}{18} u v + 
\case{3 - 3M -N + MN}{24} v^2 
+ ... \nonumber
\end{eqnarray}
The values of the coefficients $b^{(u)}_{ijkl}$, $b^{(v)}_{ijkl}$,
$e^{(\phi)}_{ijkl}$, $e^{(t)}_{ijkl}$, and $e^{(Qk)}_{ijkl}$ up to five loops
are reported in the file {\tt OMN-MS.TXT} attached to Ref.~\cite{epaps}.  The
meaning of the numbers reported in this file is the same as that of the
numbers appearing in file {\tt OMN-MZM.TXT}, see the end of Sec.~\ref{mzm}.
In file {\tt OMN-MS.TXT} the coefficients are given numerically for
simplicity, although we computed them exactly in terms of fractions and
$\zeta$ functions.

\section{The $MN$ model}
\label{secmn}

The so-called $mn$ model is defined by the Hamiltonian density 
\begin{eqnarray}
\mathcal{H} =&&
{1\over 2} \sum_{ai} \Big[ \sum_\mu (\partial_\mu \Phi_{ai})^2 + r \Phi_{ai}^2 
      \Big]   
+ {u_0\over 4!}  \Big( \sum_{ai} \Phi_{ai}^2\Big)^2 
+ {v_0\over 4!}   \sum_{abi} 
 \Phi_{ai}^2 \Phi^2_{bi},
\label{HMN}
\end{eqnarray}
where $\Phi_{ai}$ is a real $m\times n$ matrix, i.e., $a=1,\ldots,m$
and $i=1,\ldots,n$. 

We refer to Refs.~\cite{PV-rev,CPV-00,PV-00,PV-05}
for further details on the perturbative expansions in the $mn$ model.

\subsection{The 3D massive zero-momentum perturbative expansion}
\label{mzmmn}

The basic relations in the MZM scheme are the same as
those reported in Sec.~\ref{mzm}. Beside Eq.~(\ref{ren1}), we have
\begin{equation}
\Gamma^{(4)}_{ai,bj,ck,dl}(0) = 
Z_\phi^{-2} m \left( u S_{ai,bj,ck,dl} +  v C_{ai,bj,ck,dl} \right)
\label{ren2mn}  
\end{equation}
where $S$ and $C$ are the tensorial factors corresponding to the
quartic terms of the $mn$ Hamiltonian, i.e.
\begin{eqnarray}
S_{ai,bj,ck,dl} =&& \case{1}{3} 
(\delta_{ai,bj}\delta_{ck,dl} + \delta_{ai,ck}\delta_{bj,dl} 
+ \delta_{ai,dl}\delta_{bj,ck} ), \label{tensorsmn}
\\
C_{ai,bj,ck,dl} =&& 
\delta_{ij}\delta_{ik}\delta_{il}\,\case{1}{3} 
\left(\delta_{ab}\delta_{cd} + \delta_{ac}\delta_{bd} + 
      \delta_{ad}\delta_{bc} \right).
\nonumber
\end{eqnarray}
The function $Z_t$ is defined as in Eq.~(\ref{ren3}).  The
$\beta$-functions $\beta_{u,v}$ and 
RG functions $\eta_{\phi,t}$ are defined as in
Eqs.~(\ref{betamzm}) and (\ref{etaphitmzm}).
Their expansions are [we set $A=3/(16\pi)$]:
\begin{eqnarray}
\beta_{u}  =  \sum_{ijkl} b^{(u)}_{ijkl} m^k n^l u^i v^j = 
&&- u +  \case{8 + mn}{9} A u^2 
+ \case{4+2m}{9} A u v 
- \case{760  + 164 mn}{2187 } A^2 u^3  
\\&& -\case{800 + 400m}{2187} A^2 u^2 v
-\case{184 + 92m}{2187} A^2 u v^2 
+...
\nonumber\\
\beta_{v}  =
 \sum_{ijkl} b^{(v)}_{ijkl} m^k n^l u^i v^j =
&&-v + \case{8+m}{9} A v^2 + \case{4}{3} A u v   -
\case{1480 + 92 mn}{2187} A^2 u^2 v 
- \case{2096 + 400 m}{2187 } A^2 u v^2
\nonumber\\&&
- \case{760 + 164 m}{2187} A^2 v^3 + ...
\nonumber \\
\eta_\phi = \sum_{ijkl} e^{(\phi)}_{ijkl} m^k n^l u^i v^j 
= 
&&\case{16 + 8 mn}{2187} A^2 u^2 
+ \case{32+16m}{2187} A^2 u v 
+\case{16+8m}{2187} A^2 v^2 + ...
\nonumber\\
\eta_t = \sum_{ijkl} e^{(t)}_{ijkl} m^k n^l u^i v^j 
= 
&&-\case{2+mn}{9} A u  - \case{2+m}{9} A v  
+ \case{4 +2 mn}{81} A^2 u^2 
+\case{8+4m}{81} A^2 u v 
+\case{4+2m}{81} A^2 v^2 + ...
\nonumber
\end{eqnarray}
The values of the coefficients $b^{(u)}_{ijkl}$, $b^{(v)}_{ijkl}$,
$e^{(\phi)}_{ijkl}$, and $e^{(t)}_{ijkl}$ up to six loops are reported in the
file {\tt MN-MZM.TXT} attached to Ref.~\cite{epaps}.  Each line of this file
contains 7 numbers.  The first one indicates the quantity at hand: 1,2,3,4
correspond respectively to $\beta_u$, $\beta_v$, $\eta_\phi$, and $\eta_t$;
the second integer number gives the number of loops; the subsequent four
integer numbers are the indices $i,j,k,l$; finally, the last real number is
the value of the coefficient.

\subsection{The minimal-subtraction perturbative expansion}
\label{msbarmn}

The RG functions in the ${\overline {\rm MS}}$ scheme are 
\begin{eqnarray}
B_u = 
\sum_{ijkl} b^{(u)}_{ijkl} m^k n^l u^i v^j=
&&\case{8+mn}{6} u^2 + \case{2+m}{3} uv 
-\case{14+3mn}{12} u^3 
- \case{22+11m}{18} u^2 v  
-\case{10+5m}{36} u v^2 + ...
\\
B_v =  \sum_{ijkl} b^{(v)}_{ijkl} m^k n^l u^i v^j = 
&&2 u v 
+ \case{8+m}{6} v^2 - \case{82+ 5 mn}{36} u^2 v 
-  \case{58 + 11 m}{18} u v^2
- \case{14 + 3m}{12} v^3
+... \nonumber\\
\eta_\phi = \sum_{ijkl} e^{(\phi)}_{ijkl} m^k n^l u^i v^j =
&&\case{2+mn}{72} u^2  + \case{2+m}{36} u v +\case{2+m}{72} v^2 
+...
\nonumber \\ 
\eta_t = 
\sum_{ijkl} e^{(t)}_{ijkl} m^k n^l u^i v^j = 
&&- \case{2+mn}{6} u  - \case{2+m}{6} v
+ \case{2+mn}{12} u^2 + \case{2+m}{6} u v + 
\case{2+m}{12} v^2 
+ ... \nonumber
\end{eqnarray}
The values of the coefficients $b^{(u)}_{ijkl}$, $b^{(v)}_{ijkl}$,
$e^{(\phi)}_{ijkl}$, and $e^{(t)}_{ijkl}$ up to five loops are
reported in the file {\tt MN-MS.TXT} attached to Ref.~\cite{epaps}.
The meaning of the numbers reported in this file is the same as
that of the numbers appearing in file {\tt MN-MZM.TXT},
see the end of Sec.~\ref{mzmmn}.

\section{The spin-density-wave $\Phi^4$ model}
\label{secsdw}

The spin-density-wave (SDW) $\Phi^4$ model is defined by the 
Hamiltonian density 
\begin{eqnarray}
\mathcal{H} =
&&\vert\partial_\mu \Phi_{1}\vert^2 +
\vert\partial_\mu \Phi_{2}\vert^2 
+r(\vert\Phi_{1}\vert^2+\vert\Phi_{2}\vert^2 )
+\frac{u_{1,0}}{2}(\vert\Phi_{1}\vert^4+\vert\Phi_{2}\vert^4)+
\nonumber\\
&&+\frac{u_{2,0}}{2}(\vert\Phi_{1}^2\vert^2+\vert\Phi_{2}^2\vert^2)
+w_{1,0}\vert\Phi_{1}\vert^2 \vert\Phi_{2}\vert^2
+w_{2,0}\vert\Phi_{1} \Phi_{2}\vert^2 +w_{3,0}\vert\Phi_{1}^*
\Phi_{2}\vert^2 
\label{lgwh} 
\end{eqnarray}
where $\Phi_{ai}$ is a complex $2\times N$ matrix field
($a=1,2$ and $i=1,...N$).

We refer to Refs.~\cite{DPV-06}
for further details on the perturbative expansions in this model.

\subsection{The 3D massive zero-momentum perturbative expansion}
\label{mzmsdw}

The RG functions of the SDW model in the MZM scheme are defined following the
same steps as in the cases considered in the preceding sections, see 
Sec.~\ref{mzm}.

The $\beta$-functions of the renormalized quartic couplings
$u_i,\,w_i$ corresponding to the quartic Hamiltoninan parameters
$u_{i,0},\,w_{i,0}$ can be written as
\begin{equation} 
\beta_\#= \sum_{ijklmp} b^{(\#)}_{ijklmp} N^p u_1^i u_2^j w_1^k w_3^l w_3^m 
\end{equation}
where the symbol $\#$ indicates $u_1,u_2,w_1,w_2,w_3$. At one loop we have
($A\equiv 1/(8\pi)$)
\begin{eqnarray}
\beta_{u_1} &=& -u_1+A\left[ (N+4)u_1^2+4u_1u_2+4u_2^2+N
w_1^2+w_2^2+w_3^2+2w_1 w_2+2w_1w_3 \right] +\ldots \nonumber\\
\beta_{u_2} &=&-u_2+ A\left[ 6u_1u_2+N u_2^2+2w_2 w_3\right] +\ldots
\nonumber\\
\beta_{w_1}
&=&-w_1+ A\left[ 2 w_1^2+w_2^2+w_3^2+2(N+1)u_1 w_1+4 u_2 w_1+2u_1
w_2+2u_1w_3\right] +\ldots \nonumber\\
\beta_{w_2}
&=&-w_2+ A\left[ N w_2^2+2 u_1 w_2+4u_2 w_3+4w_1 w_2+2 w_2
w_3\right]+\ldots \nonumber\\
\beta_{w_3} &=&-w_3+  A\left[ N w_3^2+2u_1w_3+4 u_2
w_2+4w_1w_3+2w_2w_3\right]+\ldots \nonumber 
\end{eqnarray}
Analogous expansions apply to $e^{(\phi)}$ and $e^{(t)}$ with 
coefficients $e^{(\phi)}_{ijklmp}$, and $e^{(t)}_{ijklmp}$.
The values of the coefficients $b^{(\#)}_{ijklmp}$, 
$e^{(\phi)}_{ijklmp}$, and $e^{(t)}_{ijklmp}$ up to six loops are
reported in the file {\tt SDW-MZM.TXT} attached to Ref.~\cite{epaps}.
The file is formatted as before.
Nine numbers appear in each line. The first one indicates the 
quantity one is considering: 1,2,3,4,5 correspond to the $\beta$
functions $\beta_{u_1}$, $\beta_{u_2}$, $\beta_{w_1}$, $\beta_{w_2}$,
and $\beta_{w_3}$; 6 and 7 to $e^{(\phi)}$ and $e^{(t)}$.
The second number gives the number of loops. The subsequent six integer
numbers correspond to $i$, $j$, $k$, $l$, $m$, $p$. Finally, the last 
number gives the coefficient.

\subsection{The minimal-subtraction perturbative expansion}
\label{msbarsdw}

The RG functions of the SDW model in the $\overline{\rm MS}$ scheme 
are defined following the
same steps as in the cases considered in the preceding sections, see
Sec.~\ref{msbar}.

The $\beta$-functions of
the renormalized quartic couplings $u_i,\,w_i$ corresponding
to the quartic Hamiltoninan parameters $u_{i,0},\,w_{i,0}$
can be written as
\begin{equation} 
\beta_\#=(d-4)\# + B_\#,\qquad B_\#=
\sum_{ijklmp} b^{(\#)}_{ijklmp} N^p u_1^i u_2^j w_1^k w_3^l w_3^m 
\end{equation}
where the symbol $\#$ represents $u_1,u_2,w_1,w_2,w_3$. 
At one loop we have
\begin{eqnarray}
&&B_{u_1} =(N+4)u_1^2+4u_1 u_2+4u_2^2+N w_1^2+w_2^2+w_3^2+2w_1w_2+2w_1w_3 \\
&&B_{u_2} = 6u_1u_2+N u_2^2+2 w_2 w_3 \nonumber\\
&&B_{w_1} 
=2 w_1^2+w_2^2+w_3^2+2(N+1)u_1 w_1+4 u_2 w_1+2u_1 w_2+2u_1w_3 \nonumber\\
&&B_{w_2} 
=N w_2^2+2 u_1 w_2+4u_2 w_3+4w_1 w_2+2 w_2 w_3 \nonumber\\
&&B_{w_3} = N w_3^2+2u_1w_3+4 u_2 w_2+4w_1w_3+2w_2w_3 \nonumber
\end{eqnarray}
The functions $e^{(\phi)}$ and $e^{(t)}$ have analogous expansions with 
coefficients $e^{(\phi)}_{ijklmp}$, and $e^{(t)}_{ijklmp}$.
The values of the coefficients $b^{(\#)}_{ijklmp}$, 
$e^{(\phi)}_{ijklmp}$, and $e^{(t)}_{ijklmp}$ up to five loops are
reported in the file {\tt SDW-MS.TXT}
attached to Ref.~\cite{epaps}.
The meaning of the numbers reported in this file is the same as
that of the numbers appearing in file {\tt SDW-MZM.TXT},
see the end of Sec.~\ref{mzmsdw}.

\subsection{Perturbations of the O(4)$\otimes {\rm O}(N)$ fixed points}

We report the perturbative series of the RG eigenvalues
$\Omega_1$ and $\Omega_2$ defined in App. B of the paper. 
We write 
\begin{equation}
\Omega_\#  = \sum_{ijk} \Omega^\#_{ijk} N^k u_1^i u_2^j.
\label{Omega1}
\end{equation}
The values of the coefficients $\Omega^\#_{ijk}$ are reported in the 
file {\tt Perturbations-O4ON.TXT} attached to Ref.~\cite{epaps}.
In each line we report 6 numbers. The first one specifies the 
quantity one is referring to: 1 corresponds to $\Omega_1$ in the 
MZM scheme; 2 corresponds to $\Omega_2$ in the
MZM scheme; 3 corresponds to $\Omega_1$ in the
$\overline{\rm MS}$ scheme; 4 corresponds to $\Omega_2$ in the
$\overline{\rm MS}$ scheme. The second number gives the number of loops,
the third, fourth, and fifth number refer to the indices $i$, $j$, and $k$ 
appearing in Eq.~(\ref{Omega1}). Finally, 
the last real number is the coefficient $\Omega^\#_{ijk}$.
For $N=3$ these series are reported in App. B.

\subsection{Perturbations of the $mn$ fixed points}

We report the perturbative series of the RG eigenvalues
$\Omega_1$ and $\Omega_2$ defined in App. C of the paper. 
We write 
\begin{equation}
\Omega_\#  = \sum_{ijk} \Omega^\#_{ijk} N^k u_1^i u_2^j\; .
\label{Omega2}
\end{equation}
The values of the coefficients $\Omega^\#_{ijk}$ are reported in the 
file {\tt Perturbations-MN.TXT} attached to Ref.~\cite{epaps}.
In each line we report 6 numbers. The first one specifies the 
quantity one is referring to: 1 corresponds to $\Omega_1$ in the 
MZM scheme; 2 corresponds to $\Omega_2$ in the
MZM scheme; 3 corresponds to $\Omega_1$ in the
$\overline{\rm MS}$ scheme; 4 corresponds to $\Omega_2$ in the
$\overline{\rm MS}$ scheme. The second number gives the number of loops,
the third, fourth, and fifth number refer to the indices $i$, $j$, and $k$ 
appearing in Eq.~\ref{Omega2}. Finally, the last number is the coefficient 
$\Omega^\#_{ijk}$.

\end{document}